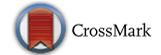 Earth, Planets and Space

**FULL PAPER**

**Open Access**

# Sunlight effects on the 3D polar current system determined from low Earth orbit measurements

Karl M. Laundal[1,2*] , Christopher C. Finlay[3] and Nils Olsen[3]

## Abstract

Interaction between the solar wind and the Earth's magnetosphere is associated with large-scale currents in the ionosphere at polar latitudes that flow along magnetic field lines (Birkeland currents) and horizontally. These current systems are tightly linked, but their global behaviors are rarely analyzed together. In this paper, we present estimates of the average global Birkeland currents and horizontal ionospheric currents from the same set of magnetic field measurements. The magnetic field measurements, from the low Earth orbiting *Swarm* and CHAMP satellites, are used to co-estimate poloidal and toroidal parts of the magnetic disturbance field, represented in magnetic apex coordinates. The use of apex coordinates reduces effects of longitudinal and hemispheric variations in the Earth's main field. We present global currents from both hemispheres during different sunlight conditions. The results show that the Birkeland currents vary with the conductivity, which depends most strongly on solar EUV emissions on the dayside and on particle precipitation at pre-midnight magnetic local times. In sunlight, the horizontal equivalent current flows in two cells, resembling an opposite ionospheric convection pattern, which implies that it is dominated by Hall currents. By combining the Birkeland current maps and the equivalent current, we are able to calculate the total horizontal current, without any assumptions about the conductivity. We show that the total horizontal current is close to zero in the polar cap when it is dark. That implies that the equivalent current, which is sensed by ground magnetometers, is largely canceled by the horizontal closure of the Birkeland currents.

## Introduction

The ionospheric current system can be decomposed into horizontal and field-aligned (Birkeland) currents. The former is often further decomposed into Hall and Pedersen currents, which are defined relative to the convection electric field in the ionosphere. It is not possible to separate these current components using only ground magnetometers. Instead, one can estimate an equivalent current, which is the two-dimensional current, flowing at some fixed height, which corresponds to the observed magnetic field disturbances (e.g., Chapman and Bartels 1940; Amm 1997). On ground at high latitudes, where the field lines are almost radial, this current can be interpreted as the divergence-free component of the horizontal current. The magnetic fields associated with curl-free horizontal currents and the Birkeland currents cancel (Fukushima 1994; Vasyliunas 2007, and references therein).

The global average high-latitude equivalent current during active geomagnetic conditions flows in two cells, which meet in the polar cap (Vestine et al. 1947; Friis-Christensen and Wilhjelm 1975; Weimer et al. 2010; Laundal et al. 2016). The pattern resembles reverse ionospheric convection during similar conditions, with some clear differences: The current in the polar cap is on average much more tilted, toward dawn on the dayside, with respect to the noon–midnight meridian than the convection streamlines. The current cell at dawn is also stronger compared to the dusk cell than what is observed in the convection. Vasyliunas (1970) explained this difference as

*Correspondence: karl.laundal@ift.uib.no
[1] Birkeland Centre for Space Science, University of Bergen, Allegt. 55, Bergen, Norway
Full list of author information is available at the end of the article a mismatch between the Hall current and the observed divergence-free currents. Friis-Christensen and Wilhjelm (1975) showed that the equivalent current is more similar to the convection pattern during summer conditions, and less so during winter.

From satellites at sufficiently low altitudes, the associated magnetic field is readily detectable. It was magnetic field measurements from ∼800 km altitude, using the TRIAD satellite, that enabled Iijima and Potemra (1978) to reveal the large-scale Region-1 (R1) and Region-2 (R2) Birkeland current systems, which is prevalent during geomagnetic active conditions. The R1 currents are located close to the polar cap boundary, the regions at polar latitudes which are threaded by field lines connected to the interplanetary magnetic field (IMF). They flow upward at the dusk side of the polar cap and downward at dawn. The R2 currents are located equatorward of the R1 currents and have opposite polarities. The large-scale Birkeland current system has later been measured and characterized by several authors. Friis-Christensen et al. (1984) used ground magnetic field measurements in combination with assumptions on ionospheric conductivity (Kamide et al. 1981) to calculate maps of the high-latitude electric field and Birkeland currents. Papitashvili et al. (2002) and Weimer (2001) used satellite data, from the Ørsted and Magsat satellites and from Dynamics Explorer 2, respectively, to develop spherical harmonic models of the Birkeland currents, resolving variations with seasons and the orientation of the IMF. He et al. (2012) used CHAMP data to develop a global model of Birkeland currents using empirical orthogonal functions. Juusola et al. (2014) also used CHAMP measurements to study global Birkeland currents, with the spherical elementary current technique (Amm 1997). With the use of relatively crude but numerous measurements of the magnetic field from the constellation of ≈70 commercial Iridium satellites, global maps of the Birkeland currents are now available at high cadence (10 min), through the Active Magnetosphere and Planetary Electrodynamics Response Experiment (AMPERE; Anderson et al. 2000; Waters et al. 2001).

Laundal et al. (2015) used the magnetic field and currents from AMPERE to compare observations in space with equivalent currents measured with ground magnetometers. They found that sunlight conditions strongly affect the equivalent current and how it relates to the Birkeland current. In darkness in the polar cap, the equivalent current measured from ground tends to align with the negative ionospheric closure of the Birkeland currents, consistent with the actual horizontal current there being close to zero. In sunlight, the equivalent current tends to align with the Hall current, and the disturbance field on ground is perpendicular to that associated with Birkeland currents in space.

In this study, we present global statistical patterns of both the horizontal equivalent current and the Birkeland currents, estimated using magnetic field measurements from the CHAMP and *Swarm* satellites, without any assumptions about conductivity. The calculations of the equivalent currents and Birkeland currents are based on a decomposition of the field into poloidal and toroidal components (Backus 1986; Olsen 1997; Weimer 2001). We use the magnetic apex coordinates systems defined by Richmond (1995), which are non-orthogonal since they take the non-dipole terms in the terrestrial magnetic field into account. The currents are well organized in these coordinates, and therefore, fewer parameters are required to describe them. The results also become more invariant with respect to spatial (longitudinal and hemispheric) and temporal variations in the Earth's field, which means that observed hemispheric and longitudinal variations can be interpreted independently of local field structures (e.g., Laundal and Gjerloev 2014).

The technique is presented in detail in the next section. "Field and current variations with sunlight conditions" section contains current patterns at high latitudes, sorted according to sunlight conditions during relatively strong solar wind driving. The results are discussed in "Discussion" section, and "Conclusions" section concludes the paper.

## Technique

We use vector magnetic field data collected by low Earth orbit satellites from the *Swarm* constellation (November 2013–September 2015) and from CHAMP (August 2000–September 2010), sampling one vector field datum every 30 s. The latest update of the CHAOS geomagnetic field model (Finlay et al. 2015), CHAOS-5, is used to obtain estimates of the time-dependent field originating in Earth's core, the static lithospheric field and the large-scale magnetospheric field, which are subtracted from the observations leaving a residual that includes the signature of the polar current systems of interest. The vector field magnetic data were rotated from the magnetometer frame to the geographic frame using Euler angles co-estimated during the construction of the CHAOS-5 field model.

The magnetic field perturbation vectors used in this study were measured at radii between $r = 6615$ km and $r = 6902$ km (geodetic heights between 248 and 546 km). At these altitudes, the satellites fly above most of the horizontal ionospheric currents, below the magnetospheric currents, and through the currents connecting the magnetosphere and the ionosphere, the Birkeland currents. Since the current density $\mathbf{J}$ does not vanish, the magnetic field is not derivable from a Laplacian potential. It can, however, be expressed in terms of two scalar potentials, $T$ and $P$, corresponding to a poloidal and toroidal component (Backus [1986]; Olsen [1997]; Sabaka et al. [2010]):

$$\Delta \mathbf{B} = \Delta \mathbf{B}^{\text{pol}} + \Delta \mathbf{B}^{\text{tor}} = \nabla \times \nabla \times \mathbf{r} P + \mathbf{r} \times \nabla T, \quad (1)$$

where $\Delta \mathbf{B}$ is the magnetic field perturbation. Since $\Delta \mathbf{B}$ is sampled in a spherical shell, which is thin compared to its radius, the poloidal field can be approximated by a Laplacian potential field, $-\nabla V$ (Backus [1986]; Olsen [1997]; Sabaka et al. [2010]):

$$\Delta \mathbf{B} = \Delta \mathbf{B}^{\text{pol}} + \Delta \mathbf{B}^{\text{tor}} = -\nabla V + \mathbf{r} \times \nabla T. \quad (2)$$

The currents associated with $\Delta \mathbf{B}^{\text{tor}}$ are radial currents flowing through the shell, i.e., the radial component of the Birkeland currents. $\Delta \mathbf{B}^{\text{pol}}$ is associated with currents that flow entirely inside or outside the shell. In this study, we assume that the latter currents have a negligible effect on $\Delta \mathbf{B}$, since the magnetospheric field is at least partly removed by use of the CHAOS model, and the remainder is usually of much smaller amplitude than the fields of interest here. $\Delta \mathbf{B}$ is thus assumed to be associated with ionospheric currents and the currents that they induce in the ground.

Equation 2 is without reference to a coordinate system. In orthogonal spherical coordinates, the potentials can be expressed in terms of spherical harmonics and related to $\Delta \mathbf{B}$ as shown by Olsen ([1997]). In the non-orthogonal magnetic apex coordinate systems (Richmond [1995]), which we use here, the standard operators in spherical coordinates do not apply. In the following, we formulate $\Delta \mathbf{B}$ in terms of scalar potentials in apex coordinates. The benefit of using magnetic apex coordinates is that the perturbation field is better organized by these coordinates, and presumably fewer parameters are needed to describe the fields (Sabaka et al. [2002]). In addition, the result will be more invariant with respect to spatial (longitudinal and hemispheric) and temporal variations in the Earth's main field.

Richmond ([1995]) defined two different magnetic apex coordinate systems: The modified apex (MA) coordinate system and the quasi-dipole (QD) coordinate system. We use a combination of these coordinate systems to represent the magnetic field. The angular coordinates of both systems are defined in terms of the field line apex, the highest point above the ellipsoid, of the adjacent International Geomagnetic Reference Field (IGRF) magnetic field line. The longitude, $\phi$, is the centered dipole longitude of the apex in both systems. The latitudes are defined as

$$\lambda_q = \pm \cos^{-1} \sqrt{\frac{R_E + h}{R_E + h_A}} \quad (3)$$

in QD coordinates and

$$\lambda_m = \pm \cos^{-1} \sqrt{\frac{R_E + h_R}{R_E + h_A}} \quad (4)$$

in MA coordinates. $h_A$ is the geodetic height of the field line apex, $R_E = 6371.2$ km is the mean Earth radius, and $h_R$ is a reference height which we set to 110 km. The equations correspond to a mapping along a dipole field line from the apex to a sphere of radius $R_E + h$ in QD coordinates and $R_E + h_R$ in modified apex coordinates. Since the MA latitude only varies with $h_A$ it is constant along field lines, while the QD latitude, which depends on the height $h$, is not.

The toroidal potential depends on currents flowing at the heights of the satellites. These currents are primarily Birkeland currents, which by definition map along magnetic field lines (which we assume are sufficiently well described by the IGRF model, ignoring the effect of $\Delta \mathbf{B}$ itself on the current path). It is therefore appropriate to use modified apex coordinates for the toroidal field. This can be done such that the potential $T$ is approximately constant along field lines (Matsuo et al. [2015]), as described below. The poloidal potential depends on remote (to the satellites) currents. At LEO, these are primarily horizontal ionospheric currents. Such currents are largely confined to the conducting layer of the ionosphere, with a maximum at approximately 110 km. To describe the radial dependence of the poloidal field, it is therefore necessary to use a coordinate system, which includes height, which is the case in QD coordinates.

### Toroidal field representation in modified apex coordinates

Assuming $T$ is constant along IGRF magnetic field lines, $\nabla T$ can be expressed in MA coordinates as (Richmond [1995])

$$\nabla T = \frac{\mathbf{d}_1}{(R_E + h_R) \cos \lambda_m} \frac{\partial T}{\partial \phi} \\ - \frac{\mathbf{d}_2}{(R_E + h_R) \sin I_m} \frac{\partial T}{\partial \lambda_m} \quad (5)$$

where

$$\sin I_m = \frac{2 \sin \lambda_m}{\sqrt{4 - 3 \cos \lambda_m}},$$

and $\mathbf{d}_1$ and $\mathbf{d}_2$ are MA base vectors defined by Richmond ([1995]). $\mathbf{d}_1$ and $\mathbf{d}_2$ are generally non-orthogonal, and their magnitudes decrease with altitude. They are perpendicular to the magnetic field of the IGRF. Following Matsuo et al. ([2015]), we write the toroidal field as

$$\Delta \mathbf{B}^{\text{tor}} = \mathbf{k} \times \left[ \frac{\mathbf{d}_1}{\cos \lambda_m} \frac{\partial T}{\partial \phi} - \frac{\mathbf{d}_2}{\sin I_m} \frac{\partial T}{\partial \lambda_m} \right] \quad (6)$$

and assume that $T$ is a function only of $\lambda_m$ and $\phi$. The radial variation is thus assumed to be included in the base vectors. $\mathbf{k}$ is an upward unit vector, relative to the ellipsoid. We represent $T$ in terms of spherical harmonics:

$$T(\lambda_m, \phi) = \sum_{n,m} P_n^m(\theta_m)(\psi_n^m \cos(m\phi) + \eta_n^m \sin(m\phi)). \quad (7)$$

where $\theta_m = 90° - \lambda_m$ and $P_n^m(\theta_m)$ are the Schmidt seminormalized associated Legendre functions of degree $n$ and order $m$. The sum over $n$ and $m$ is from $m = 0$ to $\min(n, M)$ and $n = 1$ to $N$. The truncation levels $N$ and $M$ will be specified below.

### Poloidal field representation in quasi-dipole coordinates

The poloidal potential, which is Laplacian since the currents associated with it do not intersect the satellite orbits, can be written in QD coordinates as (Richmond 1995)

$$\Delta \mathbf{B}^{\text{pol}} = -\nabla V$$
$$= -\frac{1}{R_E + h} \frac{1}{\cos \lambda_q} \frac{\partial V}{\partial \phi} \mathbf{f}_2 \times \mathbf{k}$$
$$\quad -\frac{1}{R_E + h} \frac{\partial V}{\partial \lambda_q} \mathbf{k} \times \mathbf{f}_1$$
$$\quad -\sqrt{F} \frac{\partial V}{\partial h} \mathbf{k}. \quad (8)$$

The first two terms are horizontal component of the gradient of $V$ in QD coordinates. The third term is based on the assumption that the vertical component of $\Delta \mathbf{B}^{\text{pol}}$ scales as the linear dimension of the horizontal current system. This scaling is contained in $F$, which is defined as the vertical component of the cross product of the QD base vectors $\mathbf{f}_1$ and $\mathbf{f}_2$ [see Richmond (1995) for definitions]. $V$ is expanded in terms of spherical harmonics treating the coordinates as orthogonal spherical, and the radius $r$ as $R_E + h$ (Chapman and Bartels 1940):

$$V(\lambda_q, \phi, h) = R_E \sum_{n,m} \left(\frac{R_E}{R_E + h}\right)^{n+1} P_n^m(\theta_q)$$
$$\cdot [g_n^m \cos(m\phi) + h_n^m \sin(m\phi)], \quad (9)$$

where $\theta_q = 90° - \lambda_q$. This potential corresponds to sources that are internal to the satellites, i.e., ionospheric horizontal currents. The field associated with external currents is partly removed with the subtraction of the CHAOS field model predictions, and the remaining field is assumed to be small. If it were to be included, $V$ would be split in two components, one internal [defined as in (9)] and one external, which would vary radially as $(R_E + h)^n$ instead of $(R_E + h)^{-(n+1)}$ (Olsen 1997).

### The total perturbation field

Equations (6) and (8) can be combined to give an expression for the total field variation, corresponding to (2). Measurements of $\Delta \mathbf{B}$ can then be used to estimate the spherical harmonic coefficients $\psi_n^m, \eta_n^m, g_n^m$ and $h_n^m$, through the following three equations, which refer to the geodetic east, north, and up directions (subscripts $e$, $n$, and $u$, respectively):

$$\Delta B_e = \frac{-d_{1,n}}{\cos \lambda_m} \frac{\partial T}{\partial \phi_{\text{MLT}}}$$
$$\quad + \frac{d_{2,n}}{\sin I_m} \frac{\partial T}{\partial \lambda_m}$$
$$\quad - \frac{f_{2,n}}{R_E + h} \frac{1}{\cos \lambda_q} \frac{\partial V}{\partial \phi_{\text{MLT}}}$$
$$\quad + \frac{f_{1,n}}{R_E + h} \frac{\partial V}{\partial \lambda_q} \quad (10)$$

$$\Delta B_n = \frac{d_{1,e}}{\cos \lambda_m} \frac{\partial T}{\partial \phi_{\text{MLT}}}$$
$$\quad - \frac{d_{2,e}}{\sin I_m} \frac{\partial T}{\partial \lambda_m}$$
$$\quad + \frac{f_{2,e}}{R_E + h} \frac{1}{\cos \lambda_q} \frac{\partial V}{\partial \phi_{\text{MLT}}}$$
$$\quad - \frac{f_{1,e}}{R_E + h} \frac{\partial V}{\partial \lambda_q} \quad (11)$$

$$\Delta B_u = -\sqrt{F} \frac{\partial V}{\partial h} \quad (12)$$

with $T$ and $V$ given by (7) and (9), respectively. We calculate the MA and QD base vectors and coordinates using software by Emmert et al. (2010). The apex longitude (which is equal in QD and MA coordinates) is replaced by the magnetic local time (MLT), $\phi_{\text{MLT}}$. This is because the magnetic disturbances that we estimate are primarily a result of interaction with the solar wind and therefore highly organized with respect to the Sun. We define MLT as (in radians)

$$\phi_{\text{MLT}} = \phi - \phi_{\text{noon}} + \pi \quad (13)$$

where $\phi_{\text{noon}}$ is defined as the apex longitude of the magnetic meridian that maps to the subsolar point at a sphere of radius $\gg 1\ R_E$. A large sphere is chosen to avoid the influence of low-latitude magnetic anomalies in determining the magnetic longitude, which most strongly faces the Sun. This is done because we are here most concerned with the magnetic field disturbances at high latitudes, which map far out in the magnetosphere where the terrestrial part of the magnetic field is mostly dipolar.

For studies of low latitudes, it might be more appropriate to define the noon meridian at lower heights.

Equations (10)–(12) relate the magnetic field perturbations $\Delta\mathbf{B}$ to $T$ and $V$ for any (dipole-dominated) configuration of the main magnetic field, through the use of MA and QD base vectors. $T$ and $V$ do not depend on the base vectors, and so they are independent of longitudinal, hemispheric, and temporal variations in the main magnetic field. $T$ and $V$ are represented in terms of spherical harmonics, which are not necessarily orthogonal basis functions in QD and MA coordinates. This means that the coefficients in Eqs. (7) and (9) cannot be determined from Eqs. (10) to (12) independently from each other and that the description of $T$ and $V$ may not be complete. However, in practice, these concerns are not a major obstacle; even in orthogonal coordinate systems the determined spherical harmonics coefficients are only independent if data are uniformly distributed on the sphere, while the truncation of the spherical harmonic series means that complete is never guaranteed. Below we also report a numerical test, showing that the magnetic energy is, as required, invariant between the coordinate system we use and orthogonal geographic coordinates. Furthermore, in discussion section we present a comparison between our technique and other ways of treating magnetic coordinates, which have been used in the literature, showing that our approach yields less noise and allows the retrieval of stronger, better defined currents. Thus, although we presently lack a formal mathematical justification for our scheme, we are convinced that the approximations involved are acceptable and that it represents an improvement over previous methods.

### Associated currents

The poloidal and toroidal magnetic potentials can be related to associated currents through $\mathbf{J} = \nabla \times \Delta\mathbf{B}/\mu_0$. The poloidal potential relates to a horizontal divergence-free current density, $\mathbf{J}_{\perp,df}$, which can be expressed in terms of a horizontal equivalent current function $\Psi$ at height $h_R$:

$$\mathbf{J}_{\perp,df} = \mathbf{k} \times \nabla\Psi \tag{14}$$

where

$$\Psi = -\frac{R_E}{\mu_0} \sum_{n,m} \frac{2n+1}{n} \left(\frac{R_E}{R_E + h_R}\right)^{n+1} P_n^m(\theta_q)$$
$$\cdot \left[g_n^m \cos m\phi_{\text{MLT}} + h_n^m \sin m\phi_{\text{MLT}}\right]. \tag{15}$$

The toroidal field relates to a vertical current $J_u$ at radius $R_E + h_R$ as

$$J_u = -\frac{1}{\mu_0(R_E + h_R)} \sum_{n,m} n(n+1)P_n^m(\theta_m)$$
$$\cdot \left[\psi_n^m \cos m\phi_{\text{MLT}} + \eta_n^m \sin m\phi_{\text{MLT}}\right]. \tag{16}$$

$J_u$ can be interpreted as the field-aligned current (Birkeland current), since the field lines are very close to vertical at polar latitudes. We therefore use the terms vertical (or radial), Birkeland, and field-aligned currents interchangeably. The estimated currents that we present below are calculated at $h_R = 110$ km, which is approximately the height where the Hall conductivity is highest.

The above equations for currents are based on the assumption that $T$ and $V$ can be treated as if MA and QD coordinates were orthogonal. Like $T$ and $V$, the currents are independent of longitudinal, hemispheric and temporal variations in the Earth's main field.

### Estimating the model coefficients

Finding $\psi_n^m, \eta_n^m, g_n^m$ and $h_n^m$ amounts to solving an overdetermined set of linear equations, based on typically $\sim10^5$ measurements, of the form

$$\mathbf{Gm} = \mathbf{d} \tag{17}$$

where $\mathbf{d}$ is a vector containing measurements of the magnetic field vector components $\Delta B_e, \Delta B_n, \Delta B_u$. $\mathbf{m}$ is the solution vector, containing the model coefficients $\psi_n^m, \eta_n^m, g_n^m$ and $h_n^m$. $\mathbf{G}$ is the matrix containing the relationship between $\mathbf{d}$ and $\mathbf{m}$, as given by Eqs. (10), (11), and (12). *Swarm* A and *Swarm* B fly side by side, and are therefore assumed not to provide independent data points. We therefore weight these data points by 0.5.

After the coefficient vector $\mathbf{m}$ has been determined, the equations are re-weighted according to the difference between modeled and measured values (residuals), $\epsilon_i$. The weight of the $i$'th data point, $w_i$, is defined as

$$w_i = \min(1.5\sigma/|\epsilon_i|, 1). \tag{18}$$

These weights, which are called Huber weights (Huber 1964), are then applied before solving the linear system again. The procedure is repeated until the model vector converges. $\sigma$ is the root-mean-square residual, also calculated iteratively with Huber weights. This procedure iteratively decreases the influence of outliers, including extreme events.

We truncate $V$ (Eq. 9) at $N = 35, M = 10$, and $T$ (Eq. 7) at $N = 60, M = 10$. This leads to a total of 1815 unknown coefficients.

### Testing model retrieval of poloidal–toroidal energy invariance

Above we argued that spherical harmonics in MA and QD coordinates can be used to represent $T$ and $V$, even though they may not be orthogonal basis functions in the MA and QD coordinate systems. In the absence of a mathematical proof, we carry out a numerical experiment to test a necessary, but not sufficient condition that our system is a valid representation of the magnetic field: The toroidal and poloidal magnetic energy should be the same, independently of each other, regardless of whether $T$ and $V$ are represented in magnetic coordinates or orthogonal spherical coordinates. To verify that this property is fulfilled, we specify a synthetic test potential in magnetic coordinates and calculate corresponding field values through Eqs. (10)–(12). These field values are then used to re-estimate magnetic potentials in geocentric coordinates, using the description in Olsen (1997).

In principle, any potential defined in apex coordinates could be used for this test. However, we expect that the toroidal field representation is less accurate at lower latitudes, and so we choose a synthetic test potential which is fairly realistic at high latitudes but close to constant (zero field perturbation) at low latitudes. To generate such a potential, we fit satellite data from periods when the interplanetary magnetic field (IMF) $B_z$ was between −4 and −5 nT, scaled by the function

$$\frac{\tanh(8|\lambda_q| - 2\pi) + 1}{2} \tag{19}$$

in order to reduce the low-latitude perturbations. This function is close to 1 at high latitudes and decreases rapidly to 0 around $|\lambda_q| \leq 45°$. For the synthetic test potential, the spherical harmonic expansions are truncated at $N_T, M_T = 50, 5$ and $N_V, M_V = 25, 5$.

The test is then performed by calculating the corresponding field values at ≈33,000 points, evenly spaced horizontally (in a geocentric frame), but with random radii between 6615 and 6902 km (the shell in which the satellites fly). A fixed time (00:00 UT at January 1, 2005) was chosen for the conversion from apex latitude/MLT to geocentric latitude and longitude. A spherical harmonic model is fitted to these vectors in geocentric coordinates, with truncation levels at $N, M = 65, 15$. A higher truncation is expected to be necessary since the field is less symmetrical in geocentric coordinates than in apex coordinates. Then, the model values are calculated at a new set of random radii. These values are used to calculate the magnetic energy in the toroidal and poloidal components combined and separately (the other coefficients set to zero). The results are given in Table 1. The energy is presented as the fraction (in energy at the grid points in the estimated model to that of the synthetic model. We see that ≈80 % of the energy is in the toroidal field. When the synthetic model contains only a poloidal (toroidal) component, the estimated toroidal (poloidal) component contains 0.0 % of the energy in the original field, and the poloidal (toroidal) component contains 99.9 %. Only 0.1 % of the energy in the synthetic model is not accounted for in the estimated model. We conclude that the apex representation presented above leads to magnetic energies, which are consistent with a description in geocentric coordinates, and that it does not introduce leakage of energy between the components that exceeds a fraction of one percent.

## Comparison to other datasets

In this section, we present estimates of the ionospheric current systems based on CHAMP and *Swarm* satellite data from periods when the IMF $B_z$ (GSM coordinates) was less than −2 nT, and compare the results to other datasets. IMF values are obtained from the OMNI dataset (1-min values) and are propagated in time to the magnetopause. Figure 1 shows the vertical and horizontal equivalent currents in both hemispheres estimated using the technique described above. The $B_z < -2$ nT selection criterion ensures relatively strong solar wind driving since the IMF has a component, which is antiparallel to the Earth's main field, which allows for reconnection on the dayside and subsequent energy transfer to magnetosphere. It was fulfilled for ≈3.2 × 10⁶ out of ≈15.2 × 10⁶ data points. The equivalent current (right panels) is displayed as a contour plot of the function $\Psi$, described in Eq. (15). 30 kA flow between each contour, and the total current flowing between the maximum and minimum $\Psi$ in the plotted region is written in the lower right corners. The vertical current is shown as a contour plot of the function $J_u$ in Eq. (16). The contour intervals are $0.1 \,\mu$ A/m², and red color indicates upward current and blue downward current.

The estimated equivalent current is largely symmetrical between hemispheres. It is similar to previously reported equivalent current patterns based on ground magnetometers (e.g., Vestine et al. 1947; Friis-Christensen and Wilhjelm 1975): a two-cell pattern with currents flowing largely sunward across the polar cap, however with a significant dawnward component. The parallel currents are also largely symmetrical between hemispheres and similar to the R1/R2 current patterns reported by Iijima and Potemra (1978) and observed later by several authors (e.g., Friis-Christensen et al. 1984; Weimer 2000; Papitashvili et al. 2002; Anderson et al. 2008; Green et al. 2009).

Figure 2 shows radial currents (left) and equivalent currents (right), based on AMPERE and SuperMAG data, respectively, at similar scales and format and with a similar selection criterion as in Fig. 1 (IMF $B_z < -2$ nT). They can be compared to the top row plots (Northern hemisphere) in Fig. 1.

The radial current density, which is an average of the global Birkeland current maps from AMPERE in the Northern hemisphere [evaluated at 110 km (Waters et al. 2001)], shows a very similar distribution of the Birkeland currents as in Fig. 1. However, the peak currents are sharper and stronger, by approximately a factor of 2,

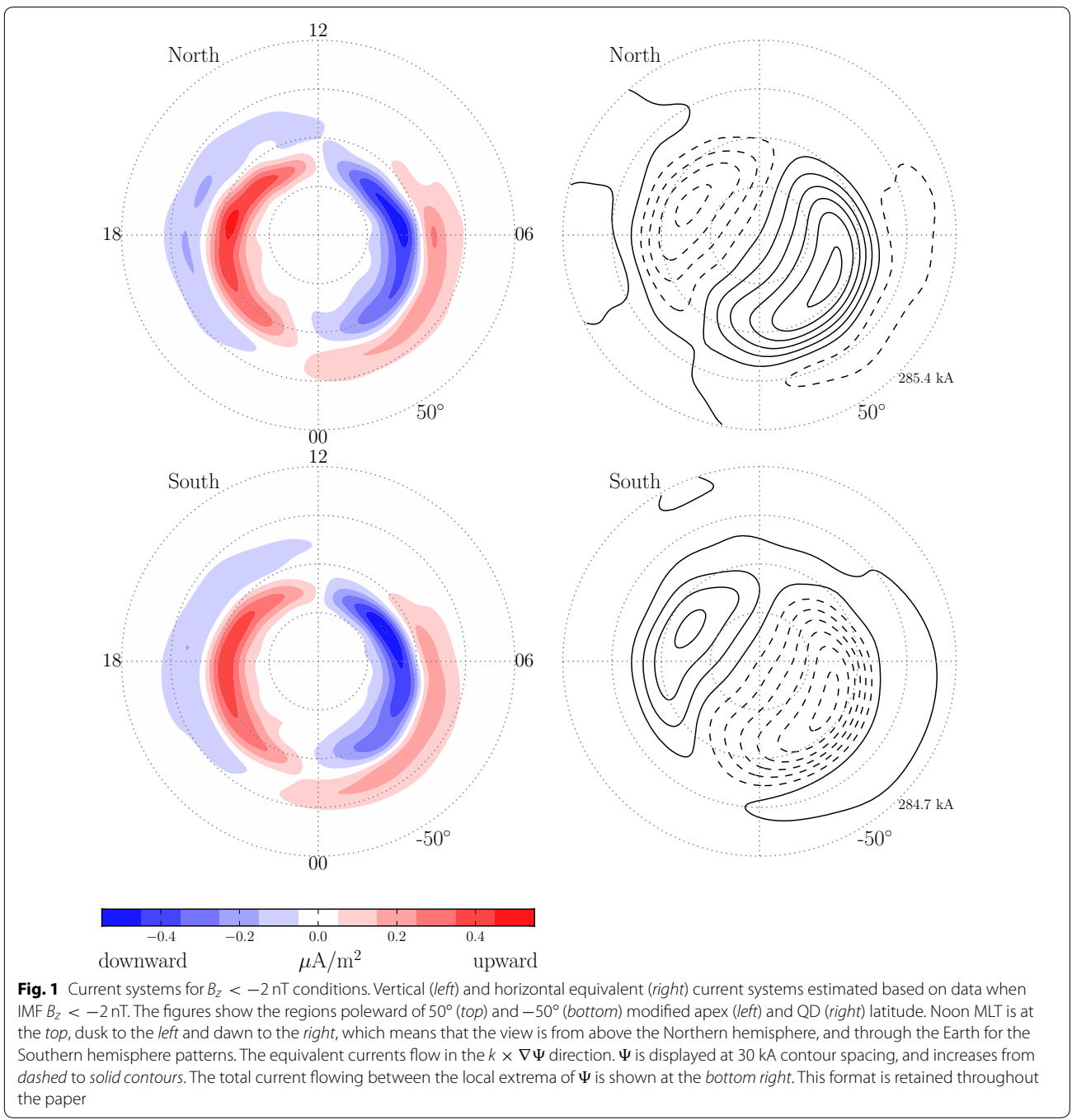

**Fig. 1** Current systems for $B_z < -2$ nT conditions. Vertical (*left*) and horizontal equivalent (*right*) current systems estimated based on data when IMF $B_z < -2$ nT. The figures show the regions poleward of 50° (*top*) and −50° (*bottom*) modified apex (*left*) and QD (*right*) latitude. Noon MLT is at the *top*, dusk to the *left* and dawn to the *right*, which means that the view is from above the Northern hemisphere, and through the Earth for the Southern hemisphere patterns. The equivalent currents flow in the $k \times \nabla\Psi$ direction. $\Psi$ is displayed at 30 kA contour spacing, and increases from *dashed* to *solid contours*. The total current flowing between the local extrema of $\Psi$ is shown at the *bottom right*. This format is retained throughout the paper

in our estimates. This may be due to the differences in how the currents are calculated; averaging global 10 min cadence current functions in the case of AMPERE, each of which are based on spherical harmonic representations of the toroidal field, and, in our technique, estimating the toroidal potential using the magnetic field values directly. Another reason for the differences may be that the quality of the CHAMP/*Swarm* magnetometers is much better than that of the Iridium magnetometers.

The equivalent current function shown in Fig. 2 is based on more than $4 \times 10^7$ SuperMAG ground magnetic field perturbation measurements at $\lambda_q \geq 49°$, from periods between 1981 and 2014 when the IMF $B_z$ was less than −2 nT. The preprocessing of the SuperMAG data

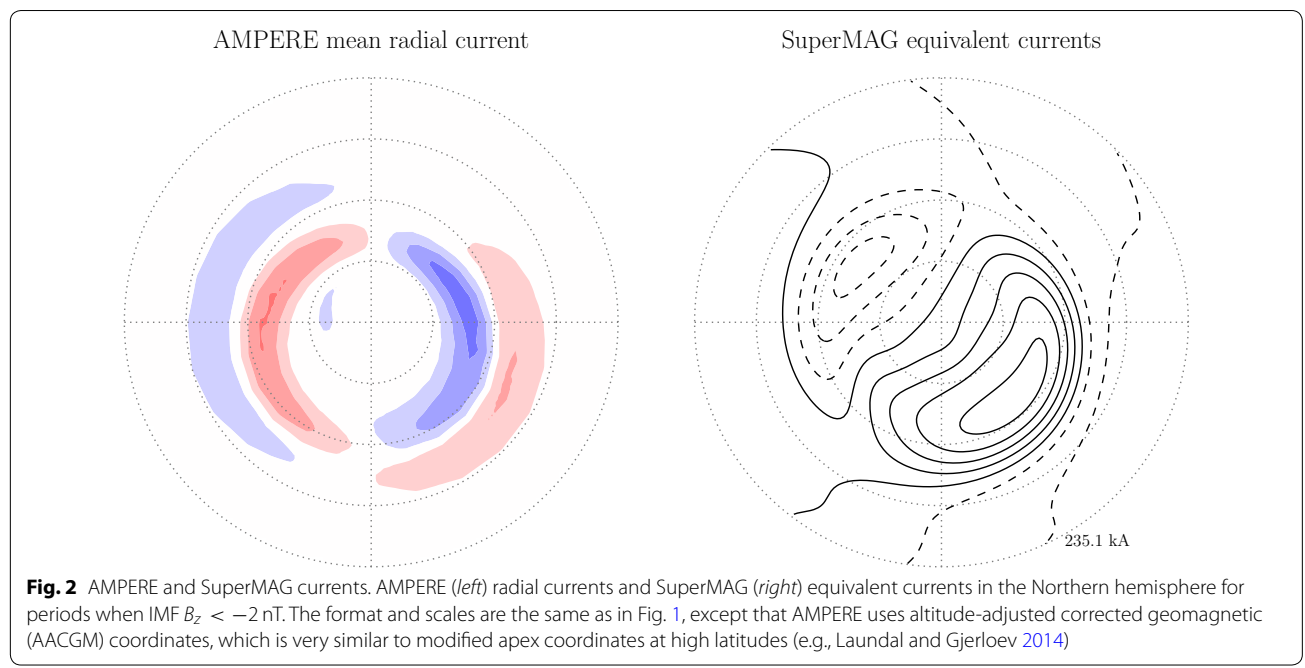

AMPERE mean radial current

SuperMAG equivalent currents

235.1 kA

**Fig. 2** AMPERE and SuperMAG currents. AMPERE (*left*) radial currents and SuperMAG (*right*) equivalent currents in the Northern hemisphere for periods when IMF $B_z < -2$ nT. The format and scales are the same as in Fig. 1, except that AMPERE uses altitude-adjusted corrected geomagnetic (AACGM) coordinates, which is very similar to modified apex coordinates at high latitudes (e.g., Laundal and Gjerloev 2014)

includes removal of an empirically determined baseline (Gjerloev 2012), which also removes diurnally recurrent patterns, presumably dominated by solar quiet currents. The magnetic field associated with these currents is present in the data from CHAMP and *Swarm*. In addition, the SuperMAG magnetic perturbation vectors, which are provided in local magnetic coordinates, have been rotated to geographic coordinates by assuming that the empirically determined horizontal field direction aligns with the IGRF model. Laundal and Gjerloev (2014) tested this assumption and found that it gave a median error of 0.2° in 106 tested magnetic observatories. Some outliers are expected to be present for magnetometers that are located close to magnetic anomalies which are not modeled in the IGRF. The equivalent current is derived from a Laplacian potential, which relates to the magnetic field perturbation as in (8). A separation of external and internal (induced) sources has been carried out, and the current is based only on the external potential. The technique, which is described in detail by Laundal et al. (2016), is equivalent to that used here, except that binned average field values are calculated prior to the inversion. In this case, 908 out of 920 bins contain data. The empty bins were located between 85° and 87° QD latitude. In the inversion, the equations (3 equations for each of the 908 bins) were weighted by the inverse of the standard error of the mean, unless this exceeded 1. Weights less than 1 occurred only in the bins at $\lambda_q > 87°$. Observations at $\lambda_q < 85°$ are therefore more strongly weighted in the SuperMAG-based currents compared to those derived from CHAMP and *Swarm*.

The resulting pattern is very similar to the corresponding Northern hemisphere pattern in Fig. 1. The main difference is an $\approx$20 % stronger total current in Fig. 1. The difference is likely underestimated, since ground-induced currents lead to underestimation of ionospheric currents from satellite heights. This difference may be an effect of the different measurement periods, different geographic coverage, different preprocessing, and/or the use of binned average vectors. Placing the equivalent current at a higher altitude would decrease the magnitude in Fig. 1 and increase the magnitude in Fig. 2.

## Field and current variations with sunlight conditions

In this section, we investigate how the current patterns change with solar illumination of the ionosphere. Figure 3 shows the estimated ionospheric currents based on data from periods when the IMF $B_z < -1$ nT, and the sunlight terminator crossed the noon meridian equatorward of 80° in the Northern hemisphere (marked by a horizontal bar and an arrow in the vertical current plot). That means that the pattern represents dark conditions in the North and sunlit conditions in the South, during relatively strong solar wind driving. The data selection used in Fig. 4 is similar except that here the Northern hemisphere is sunlit and the Southern hemisphere is dark (the sunlight terminator crossing the midnight meridian

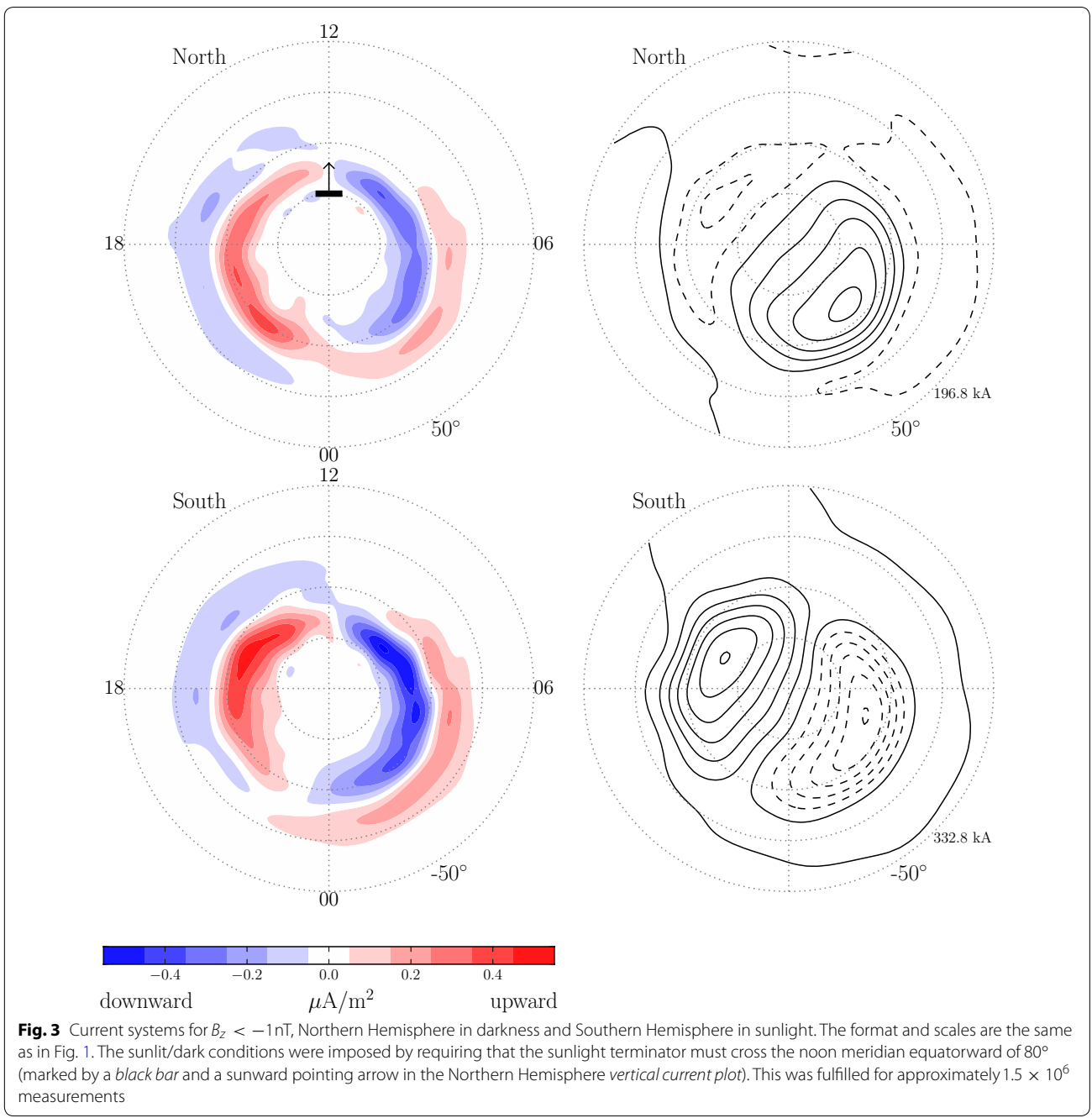

**Fig. 3** Current systems for $B_z < -1$ nT, Northern Hemisphere in darkness and Southern Hemisphere in sunlight. The format and scales are the same as in Fig. 1. The sunlit/dark conditions were imposed by requiring that the sunlight terminator must cross the noon meridian equatorward of 80° (marked by a *black bar* and a sunward pointing arrow in the Northern Hemisphere *vertical current plot*). This was fulfilled for approximately $1.5 \times 10^6$ measurements

equatorward of 80° in the North). There is no overlap in data between Figs. 3 and 4.

Figures 3 and 4 show clear differences between sunlit and dark conditions. Sunlight effects can be analyzed by comparing the Northern and Southern hemispheres in both figures separately and by comparing the same hemispheres in the two figures. The differences discussed below appear in all these combinations.

## Sunlight variations in Birkeland currents

Comparison of the Birkeland currents in the dark hemispheres with those in the sunlit hemispheres shows that on the dayside the currents are stronger in sunlight, in good agreement with previous studies (Fujii et al. 1981; Green et al. 2009; Ohtani et al. 2005a, b; Østgaard et al. 2016). The difference is most dramatic in the upward R1 current in the afternoon region. These differences can be explained in terms of differences in conductivity, which comes primarily from solar EUV emissions. However, in the pre-noon region there is on average significant diffuse electron precipitation, which seems to be relatively independent of seasons (Newell et al. 2010, Figure 9). A background of precipitation-induced conductivity will decrease the relative difference in conductance between sunlit and dark conditions, explaining the smaller dark/sunlit difference in the pre-noon currents on the dayside compared to the afternoon current.

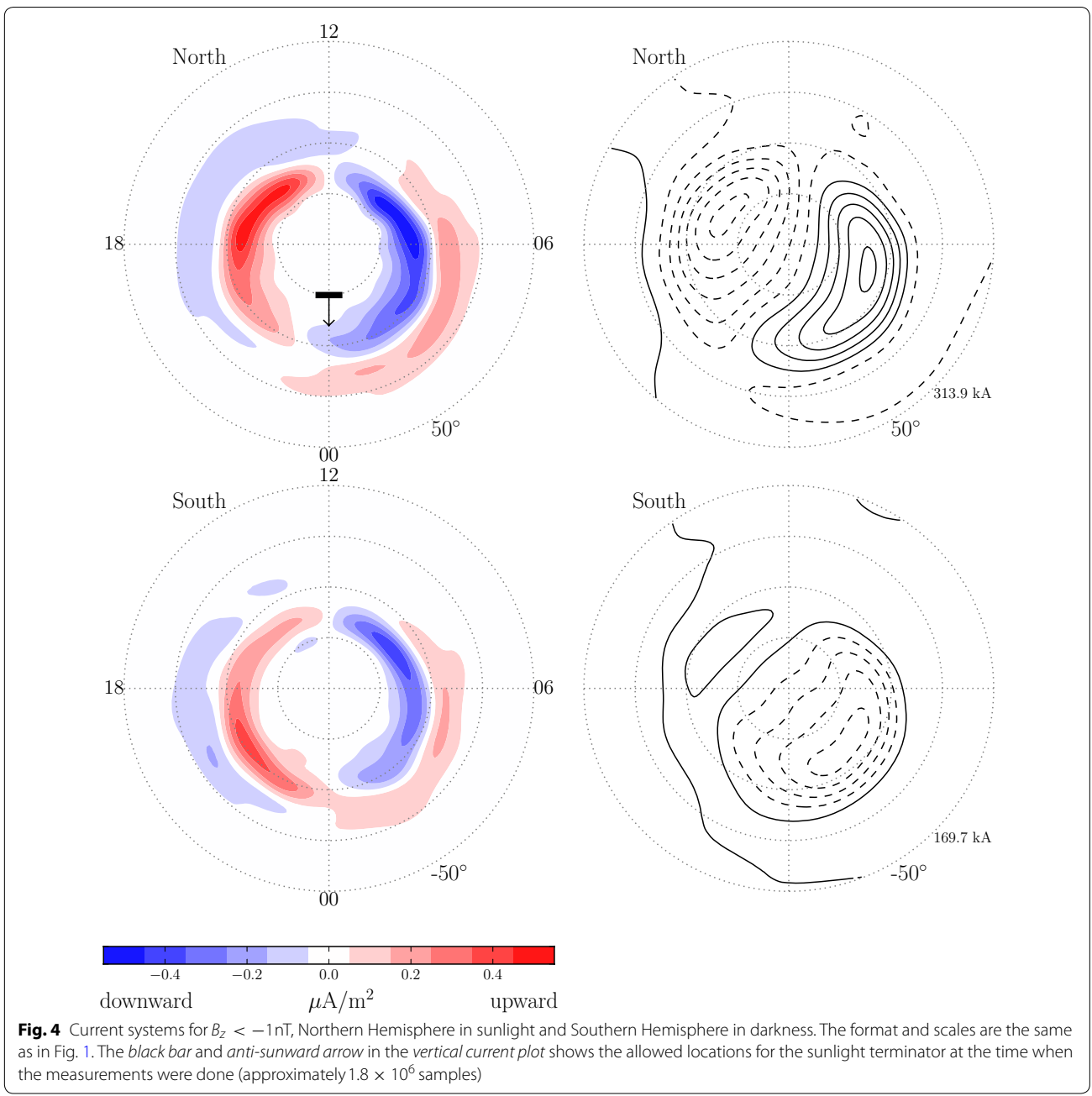

**Fig. 4** Current systems for $B_z < -1$nT, Northern Hemisphere in sunlight and Southern Hemisphere in darkness. The format and scales are the same as in Fig. 1. The *black bar* and *anti-sunward arrow* in the *vertical current plot* shows the allowed locations for the sunlight terminator at the time when the measurements were done (approximately $1.8 \times 10^6$ samples)

On the nightside, the sunlight/darkness differences have opposite polarities at dawn and dusk: In the dusk sector, the vertical currents are significantly stronger in darkness than they are in sunlight. In the dawn sector, the currents are slightly stronger in sunlight. This is also consistent with previous observations (Ohtani et al. 2005a; Green et al. 2009; Østgaard et al. 2016). The reason for the dawn/dusk difference is probably related to similar asymmetries in the precipitation of auroral particles. Newell et al. (1996, 2010) showed that electron acceleration is strongly suppressed in sunlight and that this effect is most notable in the pre-midnight sector. They argue that the reason for the suppression of electron acceleration in darkness is that field-aligned electric fields form more easily when the background conductivity is low, and/or the feedback instability (Lysak 1991), in which enhanced conductivity from precipitation increases the current, which in turn further increases the precipitation.

The R1 currents in Figs. 3 and 4 extend further poleward on the dayside when it is sunlit compared to dark conditions. This difference is seen in the entire sector between ≈5 MLT and 20 MLT. On the nightside, it is more difficult to identify a latitudinal offset because of the differences in current morphology; the dusk R1 current connecting with the dawn R2 current in darkness but not in sunlight. The difference in the dayside latitude has also been observed before, by Christiansen et al. (2002), Ohtani et al. (2005b). The latitudinal shift was explained by Ohtani et al. (2005b) as an effect of a changing magnetospheric configuration with dipole tilt angle.

### Sunlight variation in horizontal equivalent currents

The main differences between the equivalent currents in sunlight and in darkness are that (1) the overall currents are stronger in sunlight (by 69 and 85 % in Figs. 3, 4, respectively), and (2) the two current cells are of comparable magnitude in sunlight, but in darkness the dawn cell strongly dominates. The magnitude differences are not surprising since in the sunlit case there is higher ionospheric conductance from EUV illumination.

The differences in morphology are consistent with the ground-based observations by Friis-Christensen and Wilhjelm (1975) and Laundal et al. (2016). The similarities between horizontal equivalent currents derived from space and ground can be understood in terms of the Fukushima theorem (Fukushima 1994, and references therein), which states that the net magnetic effect of Birkeland currents and curl-free horizontal currents is zero on ground when the field is radial (Vasyliunas 2007). That means that the equivalent current derived from ground is the divergence-free component of the ionospheric horizontal currents. The poloidal field in space, from which the horizontal equivalent currents in the present paper are derived, is associated with currents that flow below the satellites, i.e., mainly the ionospheric currents. Only the divergence-free component of these currents has a magnetic signature (e.g., Vasyliunas 1999), and so it must be the same divergence-free current system as observed from ground.

The relationship between the divergence-free current and the Hall and Pedersen currents depends on the conductance distribution. Only if the conductances are uniform, or if their gradients are perpendicular to convection stream lines, is the divergence-free current equal to the Hall current (e.g., Laundal et al. 2015). From Figs. 3 and 4, we see that the patterns from the sunlit hemispheres resemble reverse average convection patterns for southward IMF conditions (e.g., Haaland et al. 2007; Heppner and Maynard 1987; Pettigrew et al. 2010). These are therefore most likely dominated by Hall currents, which by definition flow antiparallel to the convection streamlines. In darkness, the currents are quite different from the average convection patterns from the winter hemisphere reported by, e.g., Pettigrew et al. (2010), and thus, they must consist of more than just Hall currents. Laundal et al. (2015) showed that when the polar cap is dark, the horizontal equivalent current is antiparallel to the curl-free current, which they derived from simultaneous Birkeland current maps from AMPERE. The equivalent currents at the most polar latitudes in the dark hemispheres in Figs. 3 and 4 indeed flow in the direction connecting the peak upward and downward vertical currents (from post-midnight to pre-noon), which is consistent with the results by Laundal et al. (2015). This indicates that the actual current in the polar cap is close to zero in darkness, so that the curl-free and divergence-free (observed) currents balance.

### Sunlight-driven variations in magnetic energy density

Since we co-estimate the poloidal and toroidal potentials, we can examine the magnetic energy densities associated with each component separately. In particular, it is of interest to see how they vary with sunlight conditions, and how they relate to the energy density of the total (actual) disturbance field. Figure 5 shows the energy density calculated at $h = 400$ km in sunlit (top) and dark (bottom) conditions. The field values correspond to the Northern hemisphere currents shown in Figs. 3 and 4. The color scale is shown in units of nJ/m$^3$ (energy density) and nT (corresponding magnetic field strength). Compared to the raw data, e.g., a set of vectors $\Delta \mathbf{B}_i$ in a localized region, the energy densities in the right column of Fig. 5 are representative of the quantity $\langle \Delta \mathbf{B}_i \rangle^2 / 2\mu_0$ (where the brackets denote average). The average magnetic energy, $\langle \Delta \mathbf{B}_i^2 \rangle / 2\mu_0$ will be larger unless $\Delta \mathbf{B}_i$ are all parallel.

The energy density of the toroidal field is clearly larger than the poloidal energy density in both seasons. The distribution of the poloidal field magnetic energy density coincides with the current cells shown in Figs. 3 and 4, and the seasonal variation is also similar. The toroidal field energy density is zero at the location of the peak R1 Birkeland currents, and largest equatorward of the R1 currents. In sunlight, the toroidal field energy density has a clear sunward gradient in the polar cap, possibly an effect of the solar EUV-induced conductivity gradients. In darkness, the energy density in the polar cap is smaller and more uniform.

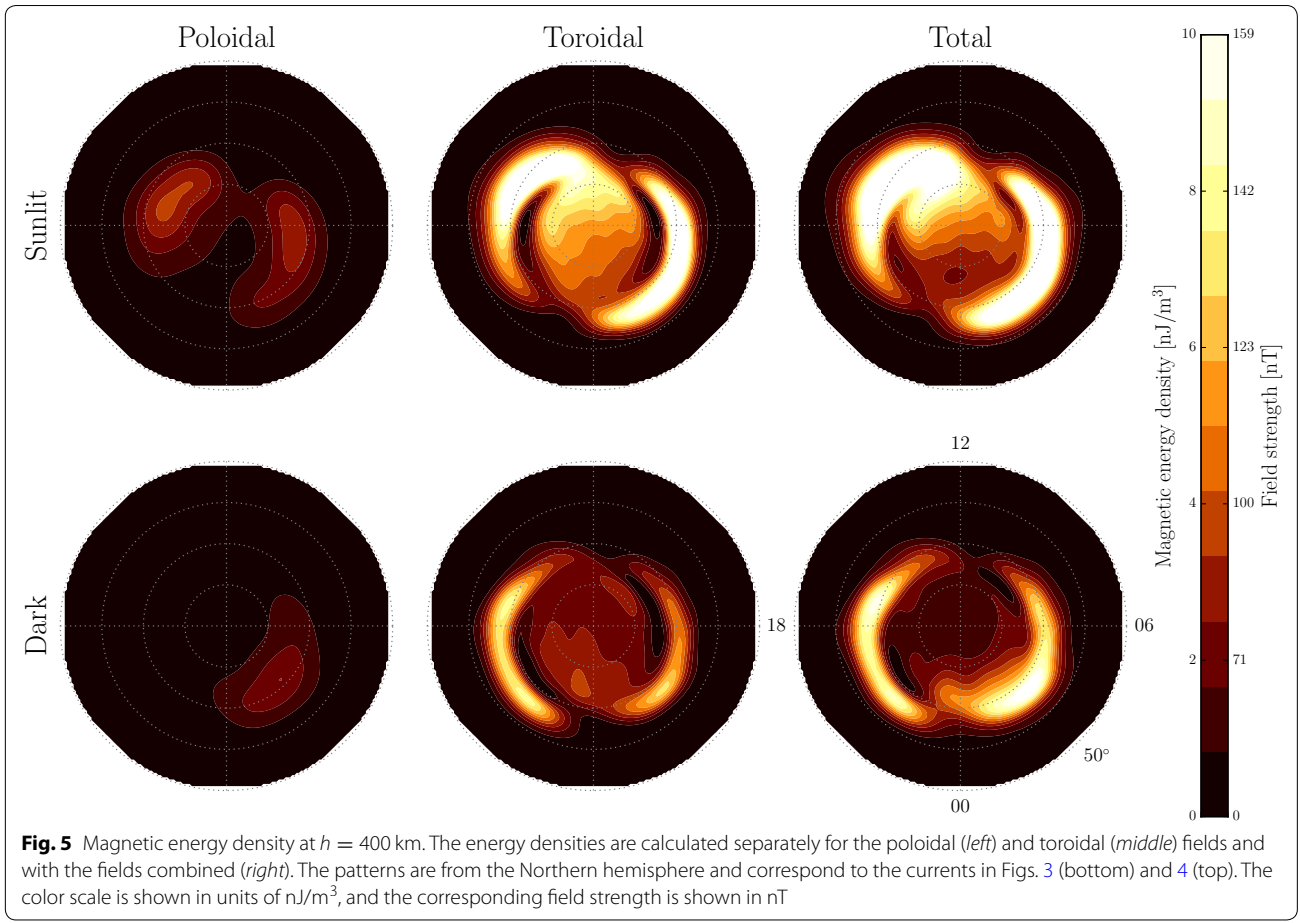

**Fig. 5** Magnetic energy density at $h = 400$ km. The energy densities are calculated separately for the poloidal (*left*) and toroidal (*middle*) fields and with the fields combined (*right*). The patterns are from the Northern hemisphere and correspond to the currents in Figs. 3 (bottom) and 4 (top). The color scale is shown in units of nJ/m³, and the corresponding field strength is shown in nT

The toroidal field energy density exceeds that of the total field in some regions. This is particularly clear in the polar caps. The reason for that is that $\Delta \mathbf{B}^{tor}$ and $\Delta \mathbf{B}^{pol}$ partly cancel, so that the vector sum is smaller than $\Delta \mathbf{B}^{tor}$ in magnitude. The angle between the two field components is shown in Fig. 6 for dark (left) and sunlit (right) conditions. We see that the components are largely antiparallel in the polar cap when it is dark and closer to perpendicular in sunlight. In the sunlit case, the angle between the components gets closer to 90° further toward the dayside. This is analogous to the results presented by Laundal et al. (2015), who compared ground-based observations with observations of the toroidal field in space, from AMPERE: They found largely perpendicular fields in sunlight and parallel fields in darkness. Since the field in the present paper is observed above the ionospheric currents, the poloidal and toroidal fields are antiparallel in darkness instead of parallel.

## Discussion

We have demonstrated a technique to estimate the global ionospheric current system at polar latitudes in the non-orthogonal magnetic apex coordinates (Richmond 1995) using magnetic field measurements from the CHAMP and *Swarm* satellites. The use of apex coordinates ensures that longitudinal and hemispheric variations in the Earth's main magnetic field, due to non-dipolar contributions, do not significantly influence the results. Presumably the estimated potentials and currents are more symmetric in apex coordinates than in, e.g., dipole coordinates, and consequently fewer parameters (in this case spherical harmonic coefficients) are needed in order to describe them. The estimated currents can be interpreted as independent of the non-dipole terms in the Earth's magnetic field. That also implies that the results should be more invariant with respect to long-term changes in the main magnetic field, compared to using dipole coordinates. The obtained currents are consistent with previous results. We have directly compared the current systems with those derived from AMPERE and from SuperMAG ground-based measurements. The comparison with AMPERE showed essentially similar morphologies, but with a sharper distribution and stronger peak currents with our technique. A comparison between the ionospheric currents derived from ground-based measurements and those from our technique also shows very similar results.

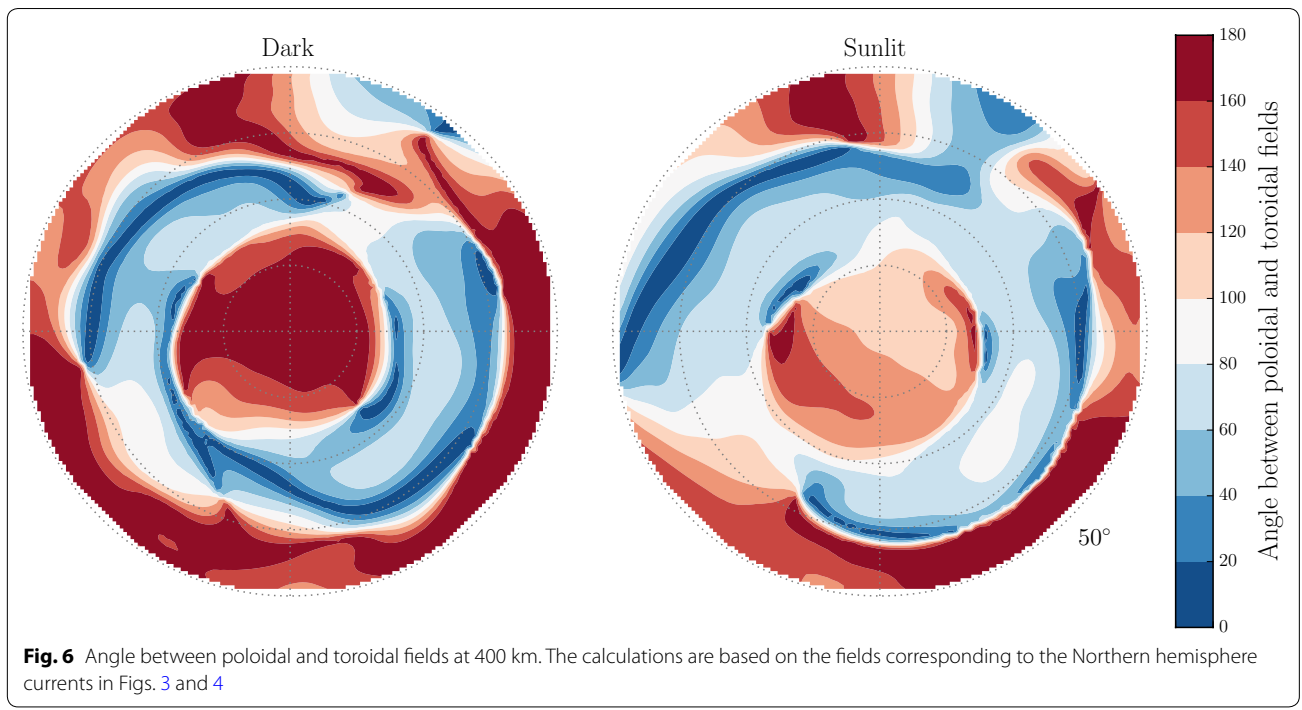

**Fig. 6** Angle between poloidal and toroidal fields at 400 km. The calculations are based on the fields corresponding to the Northern hemisphere currents in Figs. 3 and 4

The invariance with respect to longitudinal and hemispheric variations in the Earth's main magnetic field means that more reliable comparisons between hemispheres and longitudes can be carried out. Laundal and Gjerloev (2014) showed that the SML index (Newell and Gjerloev 2011), a SuperMAG-based index which is defined in a similar way as the AL index, contains variations with UT which are due to non-dipole variations in the Earth's field. This variation was reduced if the index was calculated based on the magnetic field perpendicular to QD contours of constant latitude, scaled by the distance between such contours by use of QD base vectors. The remaining variation can then be more directly interpreted in terms of conductivity variations, magnetometer distribution, and other effects, which could lead to variations with UT. Similarly, the hemispheric differences presented in this paper can be interpreted independently of the relatively large differences in the Earth's magnetic field in the two polar regions.

We specifically analyzed the difference between sunlit and dark conditions in the ionosphere, during relatively strong solar wind driving. The variations in Birkeland currents are consistent with previous results (Ohtani et al. 2005a; Green et al. 2009). The current magnitudes are largely controlled by variations in conductivity from solar EUV illumination and particle precipitation (Newell et al. 2010). We have not explicitly considered effects of IMF $B_y$ or tail dynamics, such as substorms, in this study. Comparing Figs. 3 and 4 to the study by Green et al. (2009), we see that the Birkeland current patterns resemble the situation for $B_z$ negative and $B_y = 0$. This indicates that the direct contribution of IMF $B_y$, when all values of $B_y$ are included, is on average small. We plan to investigate effects of $B_y$ and tail dynamics in more detail in future studies.

The good agreement between ground- and space-based equivalent current estimates confirms that the poloidal magnetic field in space relates to the same current system as the ground magnetic field perturbations (neglecting ground-induced currents), namely the divergence-free component of the horizontal ionospheric currents. The same conclusion was reached by Ritter et al. (2005), who compared local current systems deduced from CHAMP satellite data and the IMAGE ground magnetometer network. The horizontal equivalent current is only equal to the Hall current in special cases, depending on the conductivity and electric field in the ionosphere. The results presented here, and by Laundal et al. (2015, 2016), indicate that the Hall currents dominate the divergence-free currents in sunlit conditions, but not in darkness. The basis for this conclusion is the observation of a two-cell current pattern, similar to the average convection in reverse, in sunlight. In darkness, the dusk cell is strongly reduced, and the dawn cell dominates, which is quite different from the expected convection pattern (e.g., Pettigrew et al. 2010). One should therefore not interpret the equivalent current in darkness as the Hall current. The equivalent current is by definition a closed horizontal current, and it may be partly composed of "fictitious closure currents" (Untiedt and Baumjohann 1993) in regions where the true current is zero. Our simultaneously estimated Birkeland current patterns show that the current across the polar cap points from the peak upward to the peak downward R1 currents, consistent with it being antiparallel to the curl-free current (Laundal et al. 2015).

Satellites like *Swarm* and CHAMP, which fly relatively close to the horizontal ionospheric currents, provide the only current measurements, which can be used to derive the full current system from the same dataset. At higher altitudes, the poloidal field is greatly diminished, and only the toroidal field, associated with field-aligned currents, is observed. On ground, only the poloidal field can be observed. High-altitude data can be combined with ground-based data to estimate total current, $\mathbf{J}_\perp$ as demonstrated by Green et al. (2007). Here, we use the simultaneously estimated Birkeland current and divergence-free horizontal ionospheric current to derive the total horizontal height-integrated current. This current can be written as

$$\mathbf{J}_\perp = \mathbf{J}_{\perp,df} + \mathbf{J}_{\perp,cf} \tag{20}$$

where $\mathbf{J}_{\perp,df}$ is given by Eq. (14), and the curl-free current $\mathbf{J}_{\perp,cf} = \nabla\alpha$, where $\alpha$ relates to the Birkeland current (Eq. 16) via the current continuity equation,

$$\nabla^2\alpha = -J_u. \tag{21}$$

$\mathbf{J}_{\perp,cf}$ is thus the horizontal closure of the Birkeland currents. The solution $\alpha$ can be calculated analytically as

$$\alpha = \frac{R_E + h_R}{\mu_0} \sum_{n,m} P_n^m(\theta_m)$$
$$\cdot \left[\psi_n^m \cos m\phi_{\text{MLT}} + \eta_n^m \sin m\phi_{\text{MLT}}\right]. \tag{22}$$

This follows from Eq. (16) and the property that the spherical harmonics $Y_n^m$ satisfy

$$(R_E + h_R)^2 \nabla^2 Y_n^m = -n(n+1)Y_n^m. \tag{23}$$

Estimates of the total height-integrated ionospheric currents $\mathbf{J}_\perp$ are shown in Fig. 7. The currents in the left column are derived from the currents shown in Fig. 3 (Northern polar cap dark), and the currents in the right column are derived from the currents in Fig. 4 (Northern polar cap sunlit). The figure shows that the total current flows across the polar cap toward the afternoon region in sunlight. In darkness, the current across the polar cap is almost absent, and the current is instead largely confined to the auroral oval. Laundal et al. (2015) showed that the orientation of the curl-free and divergence-free currents is typically antiparallel in the dark polar cap, but they were not able to directly compare their magnitudes. With the present technique, we show that they do in fact approximately cancel on average.

The only other purely satellite-based estimate of the global total current pattern that we are aware of was presented by Juusola et al. (2014), who used CHAMP data and the spherical elementary current system (SECS) technique Amm (1997). A notable difference from our technique is that they used gridded geomagnetic dipole components from all heights to estimate the SECS amplitudes, on an AACGM grid. This leads to systematic errors (Laundal and Gjerloev 2014; Gasda and Richmond 1998), especially at polar latitudes, since the dipole poles and AACGM poles are at different locations. Since they mixed data from both hemispheres and did not impose any selection criteria on the data used in their total current plot, direct comparison with our results is difficult.

In Fig. 8, we present current estimates using three different combinations of magnetic coordinates, all based on the same dataset: *Swarm* A data, with $B_z < -1$ nT, and the sunlight terminator crossing the noon–midnight meridian poleward of $\pm 75°$. The field-aligned and total horizontal currents are shown. In the left column, we show estimated currents in a pure dipole coordinate representation, which is used in some studies where external and internal magnetic fields are co-estimated (e.g., Sabaka et al. 2004; Lesur et al. 2008). This is an orthogonal coordinate system, and so the mathematical treatment is exact. Nevertheless, the currents in Fig. 8 appear smeared out, because dipole coordinates do not organize magnetic disturbances sufficiently well. In the middle plot, we show the results using a combination of dipole components and apex positions, i.e., a similar approach as Juusola et al. (2014) used. Here, we see prominent features in the polar cap field-aligned currents, which are probably not realistic. These features appear because the dipole pole and apex poles are offset with respect to each other. To the right, we show estimates using the technique presented in this paper. Here the questionable polar cap features are absent. The peak currents are also stronger with a consistent treatment of apex coordinates. This leads us to conclude that our technique gives improved estimates of currents, compared to previous studies.

If conductance estimates are included, either using empirical models (e.g., Friis-Christensen et al. 1984) or auroral imagery (Lu 2013), the total current system can be estimated with ground-based data. Friis-Christensen et al. (1984) presented such estimates based on Greenland magnetometer measurements from the summer months of 1972 and 1973. Comparing their total currents for $B_z < 0$ and $B_y = 0$ to our Fig. 7 (top right) reveals that our current magnitudes are roughly 50 % of those estimated by Friis-Christensen et al. (1984). The orientation and distribution of currents are fairly similar in the auroral oval, but our currents have a much stronger duskward component in the polar cap. These differences are likely due to the conductivity model used by Friis-Christensen et al. (1984).

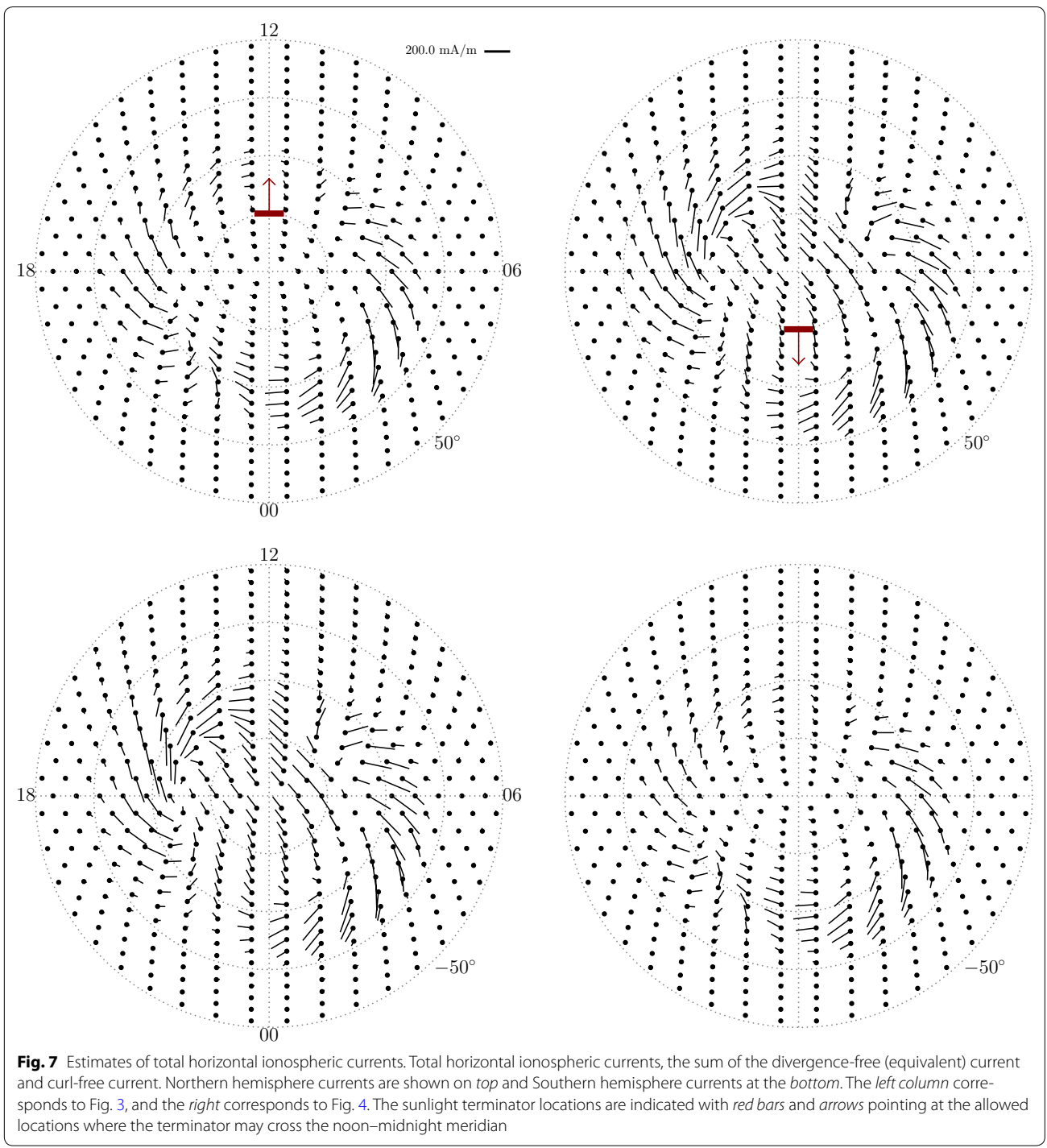

**Fig. 7** Estimates of total horizontal ionospheric currents. Total horizontal ionospheric currents, the sum of the divergence-free (equivalent) current and curl-free current. Northern hemisphere currents are shown on *top* and Southern hemisphere currents at the *bottom*. The *left column* corresponds to Fig. 3, and the *right* corresponds to Fig. 4. The sunlight terminator locations are indicated with *red bars* and *arrows* pointing at the allowed locations where the terminator may cross the noon–midnight meridian

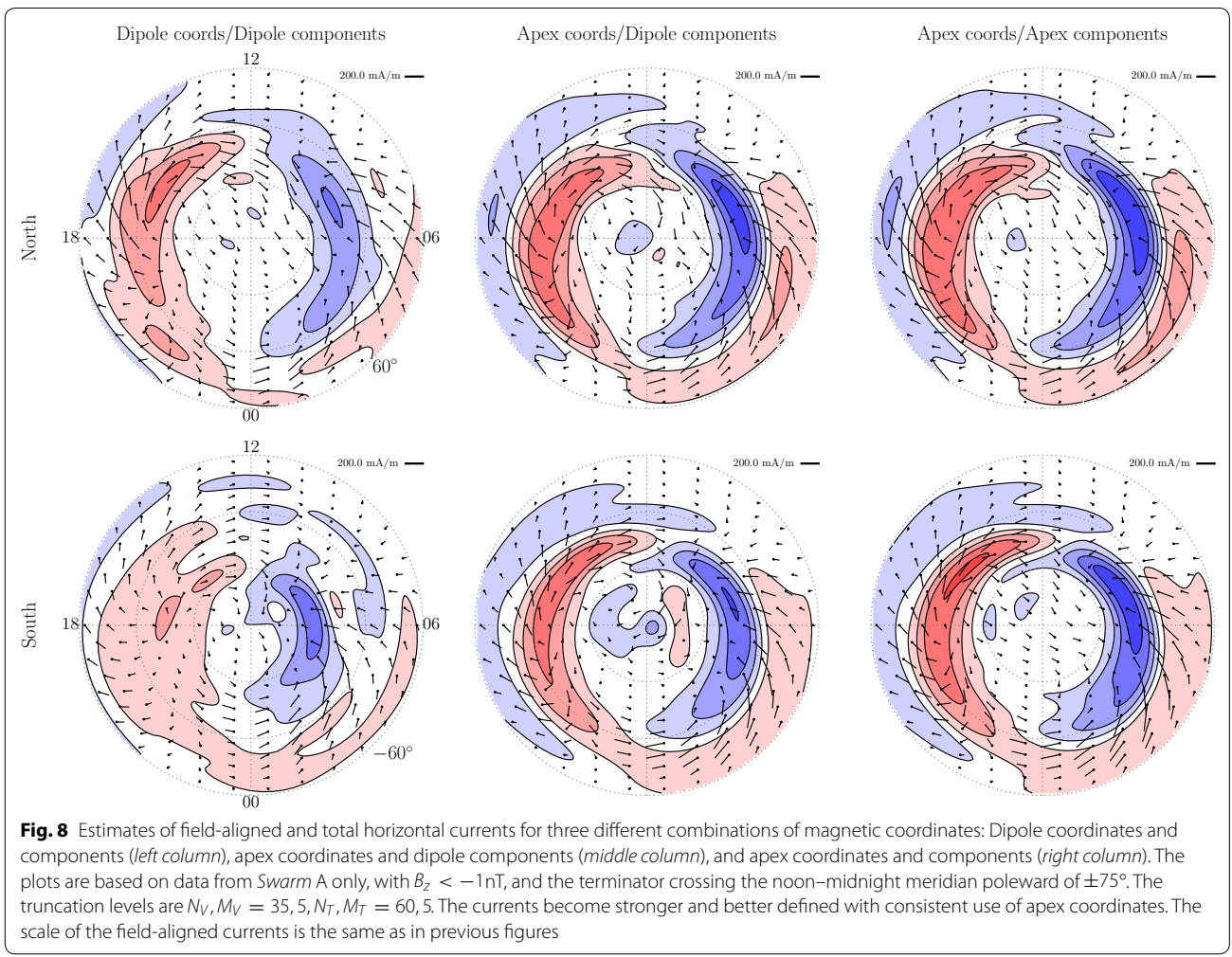

**Fig. 8** Estimates of field-aligned and total horizontal currents for three different combinations of magnetic coordinates: Dipole coordinates and components (*left column*), apex coordinates and dipole components (*middle column*), and apex coordinates and components (*right column*). The plots are based on data from *Swarm* A only, with $B_z < -1$ nT, and the terminator crossing the noon–midnight meridian poleward of $\pm 75°$. The truncation levels are $N_V, M_V = 35, 5, N_T, M_T = 60, 5$. The currents become stronger and better defined with consistent use of apex coordinates. The scale of the field-aligned currents is the same as in previous figures

**Table 1  Results of test of magnetic energy invariance between our apex description and a geocentric description**

| Estimated geocentric model components | A priori synthetic models (apex) | | |
|---|---|---|---|
| | Pol. + tor. (%) | tor. only (%) | pol. only (%) |
| *Poloidal + toroidal* | | | |
| Magnetic energy | 99.9 | 99.9 | 99.9 |
| *Poloidal only* | | | |
| Magnetic energy | 19.5 | 0.0 | 99.9 |
| *Toroidal only* | | | |
| Magnetic energy | 80.5 | 99.9 | 0.0 |

The columns represent three synthetic test models, defined in apex coordinates, with the possible combinations of toroidal and poloidal components. Field values are fitted by a model defined in geocentric coordinates, and the toroidal and poloidal components of that model are compared to the original. The numbers show the magnetic energy on a global grid as a fraction (in %) of the magnetic energy in the corresponding (apex coordinate) synthetic model. The fraction of poloidal (toroidal) energy in the synthetic model was 19.4 % (80.6 %)

Figures 3, 4 and 7 also give some insight into how the Birkeland currents close in the ionosphere. Our results show that in the case of a dark polar cap, the total current is close to zero. Several interpretations are possible as to where the Birkeland currents close in this case: A naive mathematical interpretation says that the currents close through the polar cap, through $\mathbf{J}_{\perp,cf}$, which is derived from $J_u$ and the current continuity equation. However,

$\mathbf{J}_{\perp,ef}$ is canceled by an independent opposite current, which is part of the $\mathbf{J}_{\perp,df}$ current system. A more physics-based interpretation is that the R1 currents are connected through the conducting auroral oval, by the westward currents in the post-midnight sector in Fig. 7. The continuity equation only gives information about the curl-free part of this circuit, and an arbitrary divergence-free part may be added. This divergence-free part cancels the curl-free part in the polar cap and adds to the curl-free current in the auroral oval. Since it is the divergence-free part, which is observed from ground, this interpretation is consistent with the westward electrojet being directly related to Birkeland currents in the dark hemisphere.

It is frequently stated that Birkeland currents are closed by Pedersen currents (e.g., Le et al. 2010; Ganushkina et al. 2015), although this is only true in special cases (e.g., Vanhamäki et al. 2012), depending on the conductivity. Since $\mathbf{J}_{\perp}$ is close to zero in the polar cap in darkness, the Pedersen current must also be small there. Since the electric field is typically north–south in the auroral oval, due to sunward ionospheric convection, the closure current there is probably dominated by Hall currents. Consequently, the quantity $\mathbf{E} \cdot \mathbf{J}_{\perp}$, which in the neutral wind frame of reference largely represents energy dissipation through frictional heating (Vasyliunas and Song 2005), must be much stronger on average in sunlight than it is in darkness. This is consistent with the result by Weimer (2005), who reports much larger dissipation globally in the Northern hemisphere for positive dipole tilt angles. The reduced importance of Pedersen currents in darkness implies that Birkeland current closure is partly dissipationless.

In the sunlit part of the ionosphere, the divergence-free part largely corresponds to Hall currents, since it resembles reverse convection patterns [see Figs. 3, 4, and the study by Laundal et al. (2016)]. Then the curl-free part will correspond to a true current component, the Pedersen current, which in this case contributes to Birkeland current closure. The westward electrojet will be dominated by Hall currents in sunlight. We therefore interpret the westward electrojet as being maintained by Hall currents in sunlight, and by Birkeland current closure in darkness. This explains the relatively modest differences between solstices in the post-midnight westward equivalent currents in Figs. 3 and 4.

The technique that has been presented here can also be used to analyze ionospheric currents at low latitudes (Sabaka et al. 2002), and we implicitly include effects of these currents since we use global data. However, we have made some choices to optimize the technique for use at polar latitudes: (1) The data distribution is not uniform, and it disproportionately favors polar latitudes. (2) According to Matsuo et al. (2015), the toroidal potential

$T$ (7) is approximately constant along field lines at high latitudes, but this may not be the case at lower latitudes. (3) Our definition of MLT, using the subsolar point on a sphere with very large radius, is chosen to organize polar disturbances with respect to the Sun as precisely as possible. For studies of lower latitudes, it would be more appropriate to choose a smaller sphere in order to account for non-dipole terms.

## Conclusions

We have presented a simultaneous estimation of the ionospheric equivalent horizontal current and Birkeland current on a global scale from the same set of magnetic field measurements. The estimated magnetic potentials were represented in the non-orthogonal modified magnetic apex and quasi-dipole coordinate systems, which take the full IGRF model into account. The estimated Birkeland currents are similar to, but more detailed than the average patterns derived with AMPERE. The estimated horizontal current is similar to the equivalent current derived from ground magnetometers.

An analysis of differences between sunlit and dark conditions during IMF $B_z < -1$ nT conditions shows that:

1. Birkeland currents on average scale with the conductivity, which is dominated by sunlight EUV emissions on the dayside, and by particle precipitation in the pre-midnight region. This is in agreement with the results by Ohtani et al. (2005a) and Green et al. (2009). Particle precipitation also leads to smaller differences between sunlit and dark conditions in the pre-noon sector compared to the afternoon sector.

2. The dayside (approximately 5–20 MLT) poleward boundary of the Birkeland currents is shifted poleward in sunlit conditions compared to dark. This is probably an effect of the dipole tilt angle affecting the magnetospheric configuration (Ohtani et al. 2005b).

3. The equivalent current is stronger in sunlight and resembles average two-cell convection pattern for similar conditions. In darkness, the dawn cell strongly dominates. The similarity between current and convection (only with opposite orientations) in sunlight strongly indicates that Hall currents dominate.

4. The total horizontal ionospheric current is close to zero in the dark polar cap. That implies that the divergence-free current, which is sensed by ground magnetometers, is largely canceled by the curl-free current. This is consistent with the results obtained by Laundal et al. (2015).

5. The observed currents in dark conditions are consistent with the R1 Birkeland currents being connected by a westward current through the conducting auro-ral oval. This westward current is observable from ground as the auroral electrojet. The electrojet thus relates directly to Birkeland currents in darkness, but not necessarily in sunlight.

The technique outlined here can, with some modifications discussed above, be used to analyze currents at low latitudes in quasi-dipole coordinates. We have also demonstrated that the simultaneously measured Birkeland current and divergence-free horizontal current can be used to calculate the actual horizontal (height-integrated) current, without any assumptions about conductivity. The *Swarm* satellite constellation will continue to provide measurements for several years to come. That means that the accuracy of statistical studies such as these will be improved and that further parametrization of the model parameters, e.g., in terms of the IMF and seasons, will become more realizable.

**Author details**
[1] Birkeland Centre for Space Science, University of Bergen, Allegt. 55, Bergen, Norway. [2] Teknova AS, Kristiansand, Norway. [3] DTU Space, National Space Institute, Technical University of Denmark, Kgs. Lyngby, Denmark.

**Acknowledgements**
ESA is thanked for providing prompt access to the *Swarm* L1b data. The support of the CHAMP mission by the German Aerospace Center (DLR) and the Federal Ministry of Education and Research is gratefully acknowledged. The IMF, solar wind and magnetic index data were provided through OMNIWeb by the Space Physics Data Facility (SPDF), and downloaded from ftp://spdf.gsfc.nasa.gov/pub/data/omni/highresomni/ We thank the AMPERE team and the AMPERE Science Center for providing the iridium-derived data products. The data are available at http://ampere.jhuapl.edu/ For the ground magnetometer data, we gratefully acknowledge SuperMAG, PI Jesper W. Gjerloev, and its data providers. The data are available at http://supermag.jhuapl.edu/ This study was supported by the Research Council of Norway/CoE under contract 223252/F50.